\documentclass[9pt,twocolumn,twoside]{osajnl}
\graphicspath{{.}{./figs/}}
\journal{ol} 

\setboolean{shortarticle}{true}

\title{Highly-efficient broadband TE/TM polarization beam splitter}

\author[*]{Kiyanoush Goudarzi, Doyoung Kim, and Haewook Han}

\affil[ ]{Department of Electrical Engineering, Pohang University of Science and Technology, Pohang 37673, Republic of Korea}

\affil[*]{Corresponding author: hhan@postech.ac.kr}

\begin{abstract}
Highly-efficient broadband polarization beam splitters with small footprint and based on silicon photonics technology are of great importance in optical integrated circuits. In this paper, a crucially applicable optical component of a TE/TM polarization beam splitter has been designed and simulated using the inverse design of adjoint method. The mentioned device has ultra-high transmission efficiency (about 96\% for TM\textsubscript{00}  and 93\% for TE\textsubscript{00} modes) and high extinction ratio of more than 16 dB over the bandwidth of 200 nm. Small footprint (2.2 $\times$ 2.1 \textmu m\textsuperscript{2}), broad bandwidth (200 nm), and compatibility with CMOS technology fabrication indicate the eligibility of this device.
\end{abstract}

\setboolean{displaycopyright}{true}

\begin{document}

\maketitle\\

\noindent All-dielectric silicon-based optical devices are of great interest because of their wonderful properties \cite{jahani2016all,zhao2009mie}. Some of these properties include ultra low energy dissipation, excitation of electric and magnetic dipole moments, and compatibility with CMOS fabrication technology. Unlike metallic structures with high dissipation (because of their numerous free electrons) \cite{boltasseva2011low,yao2014plasmonic,hao2011nearly,li2015circularly}, Si based optical devices have very low energy dissipation. Due to high refractive index of Si (about 3.48) for a 1.55 \textmu m communication window, many resonances of electric and magnetic dipole moments can be excited with respect to the incident electric and magnetic fields with frequency below or near Si bandgap that can be described by Mie theory \cite{zhao2009mie,butakov2016designing,muhlig2011optical,ghanekar2017mie,miroshnichenko2012optically,staude2017metamaterial}.

Because of the high refractive index of Si and high refractive index contrast between Si and SiO\textsubscript{2}, Si is applicable for high-density optical integrated circuits (OIC). Due to the high refractive index of silicon, birefringence occurs in silicon waveguides, causing polarization mode dispersion (PMD) \cite{fukuda2006ultrasmall}. One approach to overcome this difficulty is a precise fabrication process. Applying polarization beam splitters (PBSs) and rotators is another approach to eliminate the PMD effect. PBSs have other applications, e.g., polarization-based imaging systems \cite{bogaert2007led}, and magneto-optic storage heads \cite{kostuk1994diffractive}. Based on the above mentioned, designing an all-silicon PBS with high efficiency, broad bandwidth, small footprint and compatibility with CMOS fabrication technology is necessary. Better than conventional methods, an optimized design approach for desire PBSs is utilizing optimizations and artificial intelligence (machine/deep learning) methods; optimization methods are more efficient and faster than the others \cite{tahersima2019deep,ma2020inverse,ma2020arbitrary,piggott2015inverse,piggott2017fabrication}.

Few number of PBSs has been designed using the optimization methods as following. In 2015, Shen et al. proposed a PBS based on direct binary search optimization method, with digitized air holes in silicon slab on SiO\textsubscript{2} substrate; the footprint, band width, efficiency, and extinction ratio (ER) were, respectively, 2.4 $\times$ 2.4 \textmu m\textsuperscript{2}, 32 nm, 70\%, and 10 dB \cite{shen2015integrated}. A few years ago, Huang et al. fabricated an optimized PBS with adjoint inverse design based on the Si embedded in SiO\textsubscript{2}; the parameters were, correspondingly, 0.48 $\times$ 6.4 \textmu m\textsuperscript{2}, 89.9\%, 71 nm, and 14.5 dB for footprint, transmission, bandwidth and ER \cite{huang2018ultra}. There are another studies of PBSs with two dielectric waveguides structure that has been designed using particle swarm optimization (PSO) method which suffer from big structure and narrow bandwidth \cite{zhang2019particle,chen2020ultra}. In these studies, PSO was used to optimize the dielectric waveguides width for accommodating the desire outputs.

All of these PBSs suffer from low efficiency, ER, and bandwidth and high simulation time. In this letter, an on-chip TE/TM PBS, that splits input polarizations (TE\textsubscript{00}/TM\textsubscript{00}) into two fundamental TE and TM polarization modes with high efficiency, ER, and broad bandwidth, will be designed and simulated in short time using the adjoint inverse design.

The device footprint is 2.2 $\times$ 2.1 \textmu m\textsuperscript{2} consisting of Si on the buried SiO\textsubscript{2} substrate. As for 3D FDTD simulations, the mesh sizes have been chosen as $dx$ = 20 nm, $dy$ = 20 nm, and $dz$ = 50 nm, respectively in $x$-, $y$-, and $z$-directions. For designing the device, adjoint optimization method has been used \cite{lalau2013adjoint,Millerthesis}. This optimization method utilizes gradient descent algorithm to design the optimized structure. Due to the fine mesh sizes of the structure, the simulation time for designing this device is 3 h with the computational resource of a PC with 3 GHz Core-i9 CPU, and 128 GB RAM.

\begin{figure}[ht]
\centering
\includegraphics[width=\linewidth]{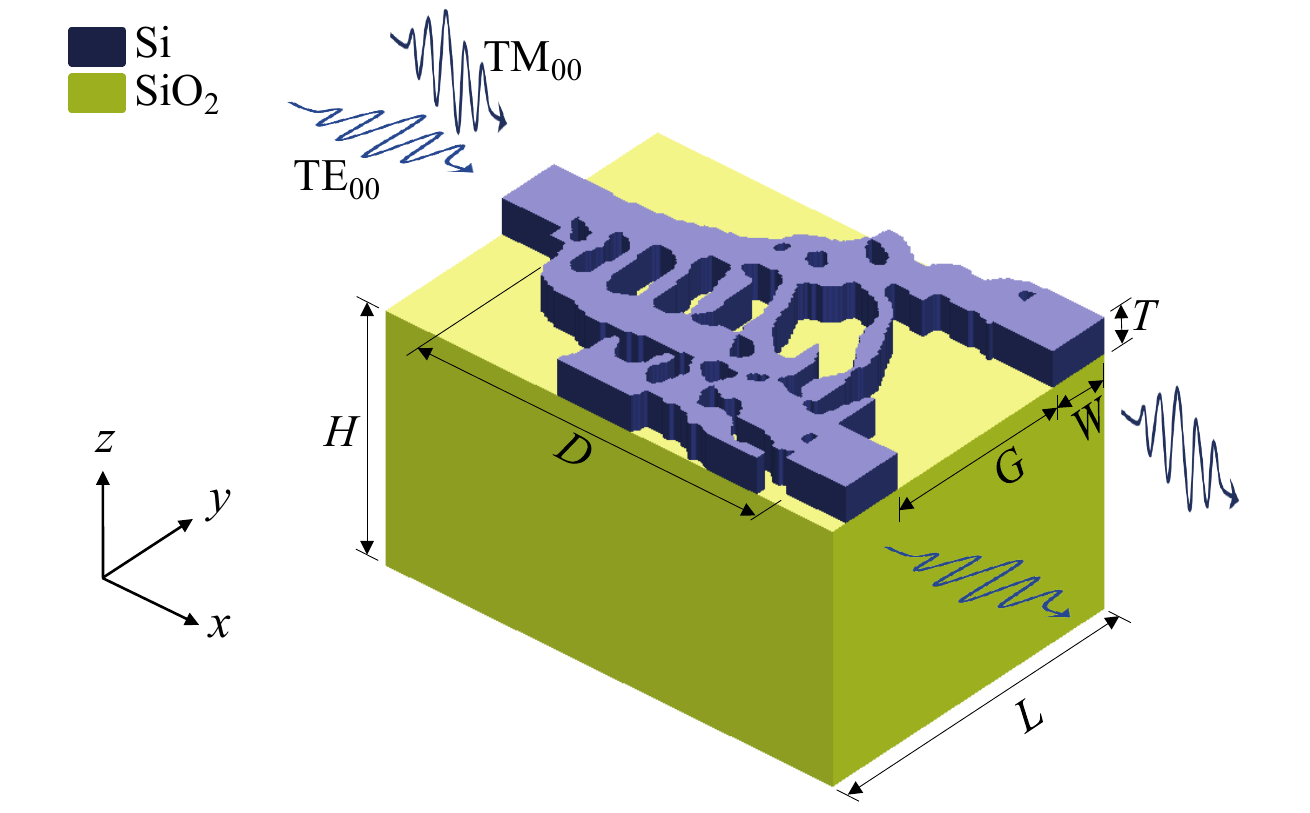}
\caption{Schematic of the optimized PBS structure.}
\label{fig:structure}
\end{figure}

\begin{figure}[h]
\centering
\includegraphics[width=\linewidth]{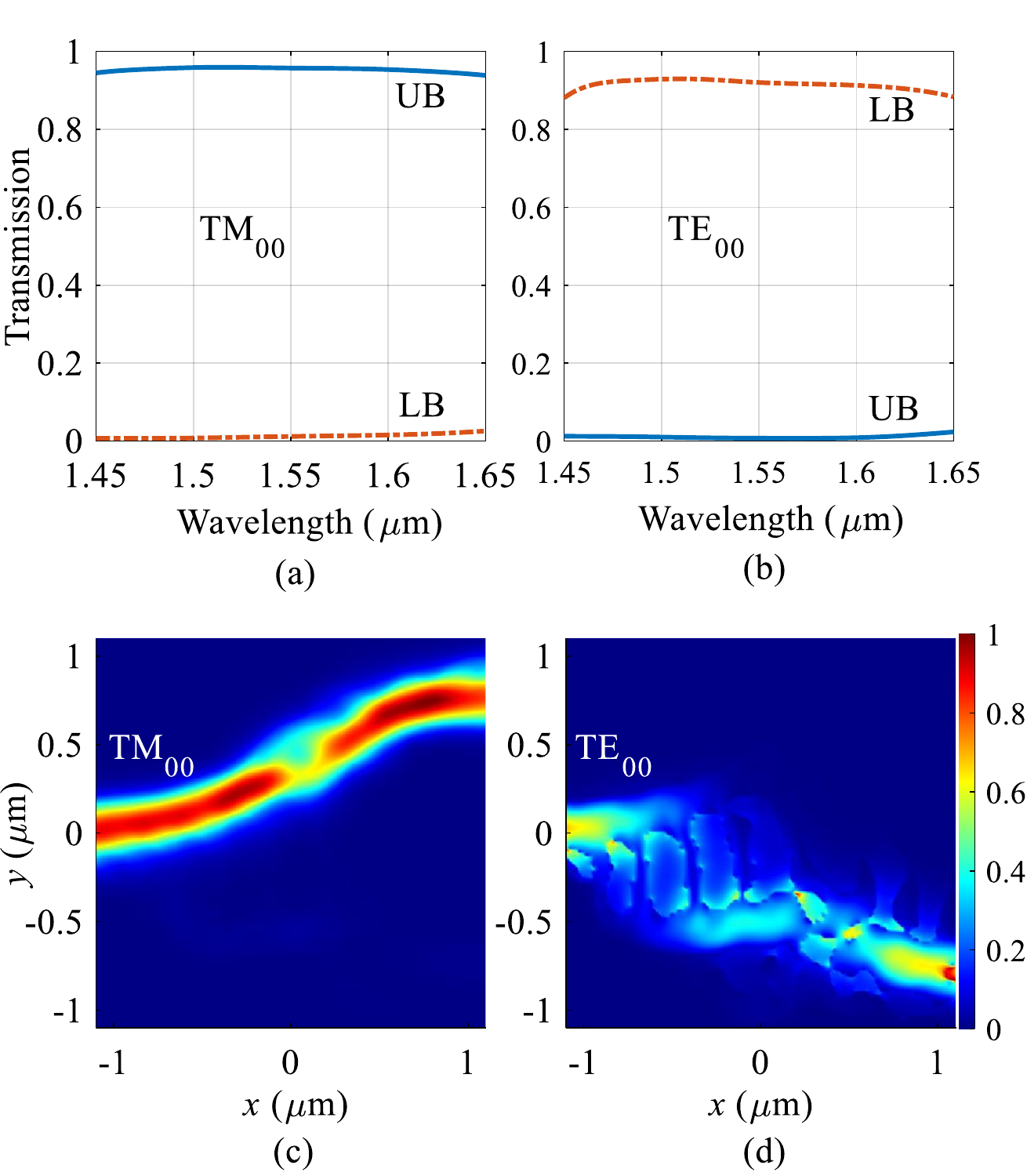}
\caption{(a) and (b) show transmission for incident waves with TM\textsubscript{00} and TE\textsubscript{00} modes, respectively. (c) and (d) depict distribution of electromagnetic power over the structure for (a) and (b), sequentially, at 1.55 \textmu m wavelength.}
\label{fig:fields}
\end{figure}

Figure \ref{fig:structure} shows the optimized structure of the proposed PBS. In this figure, $W$, $G$, $L$, $D$, $H$, and $T$ parameters are 0.4, 1.2, 2.2, 2.1, 3, and 0.4 \textmu m, respectively. Also, Si and SiO\textsubscript{2}  have the refractive indices of 3.48 and 1.44 in a 1.55 \textmu m communication window, correspondingly. Figure \ref{fig:fields} shows 3D FDTD simulation results for the proposed PBS in TE\textsubscript{00} (electrical field in $y$-direction) and TM\textsubscript{00} (electrical field in $z$-direction). In this figure, UB and LB stand for upper and lower branches of the device.

As obvious from the simulations, transmission at the upper and lower branches for the incident TM\textsubscript{00} polarization mode is about 96\% and 1.5\%, respectively (Fig. \ref{fig:fields}(a)). According to Fig. \ref{fig:fields}(b), the transmission for TE\textsubscript{00} polarization mode as the incident beam is almost 93\% at the lower and about 1.4\% at the upper branches. The electromagnetic power distributions at 1.55 \textmu m wavelength, which has been demonstrated in Figs. \ref{fig:fields}(c) and (d) for both TM\textsubscript{00} and TE\textsubscript{00} polarization modes, show high coupling between the input and the desired outputs.

Figure \ref{fig:fields}(c) and (d) indicate that there is one optical path for TM\textsubscript{00} and another path for TE\textsubscript{00} modes; the former path has nearly the same width over its length and because of about uniform waveguide width, it can only support TM\textsubscript{00} mode. Therefore, the input TM\textsubscript{00} mode can only propagate in the upper waveguide with the total internal reflection.

Further to mention, subgrating optical path is known as the propagation path for the TE\textsubscript{00} mode. This path consists of two subgrating couplers as well as two waveguides (Fig. \ref{fig:structure}). Because of the phase matching between the incident TE\textsubscript{00} light and the first grating coupler, the light can be coupled to the first waveguide. Moreover, due to another phase matching between the light and the second subgrating, it is coupled to the output waveguide. The subgratings play the role of changing Si refractive index to the desired value which is suitable for the TE\textsubscript{00} mode propagation. The TE\textsubscript{00} mode effective refractive index for a subgrating can be calculated using $n_\textrm{b}^2\approx fn_{\textrm{c}}^2 + (1-f)n_{\textrm{cl}}^2 $, where $n_\textrm{b}$, $n_\textrm{c}$, and $n_\textrm{{cl}}$ are effective refractive index of subgrating and refractive index for silicon and air in this structure; also, $f$ is fill factor of the subgrating structure, correspondingly \cite{feng2007polarization}.

\begin{figure}[h]
\centering
\includegraphics[width=\linewidth]{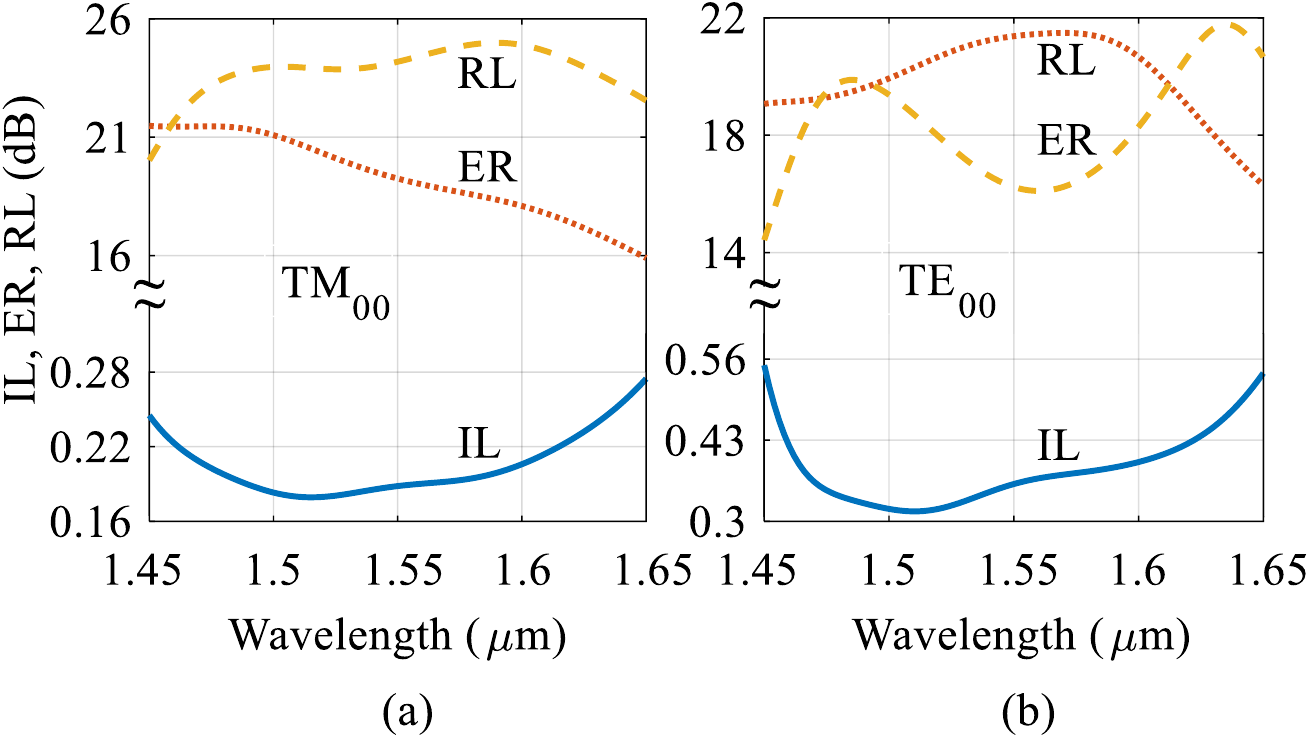}
\caption{IL, ER, and RL for (a) TM\textsubscript{00} and (b) TE\textsubscript{00} modes.}
\label{fig:dB}
\end{figure}

Furthermore, Fig. \ref{fig:dB} shows insertion loss (IL), ER and return loss (RL) for the device in the TM\textsubscript{00} and TE\textsubscript{00} modes over 200 nm bandwidth. As obvious from Figs. \ref{fig:dB}(a) and (b), IL is less than 0.28 and 0.55 dB for TM\textsubscript{00} and TE\textsubscript{00} modes, respectively. ER and RL are more than 16 and 20 dB for TM\textsubscript{00} mode; also, these parameters are more than 16 and 14 dB for TE\textsubscript{00} modes (Fig. \ref{fig:dB}).

In this letter, an all-dielectric metamaterial-based structure for realizing PBS were optimized and simulated using adjoint and 3D FDTD methods. Based on the results, the proposed TE/TM PBS has many advantages over the other TE/TM PBSs, such as high transmission efficiency (about 96\% for TM\textsubscript{00} and 93\% for TE\textsubscript{00}), high ER (> 16 dB), broad and flat bandwidth (200 nm), and short optimization time (3 h). This device is useful for isolating in optical communications, polarization-based imaging systems, and magneto-optic storage heads; furthermore, it is applicable to eliminate mode dispersion effect in birefringence waveguides. The proposed device can be consider as one of the efficient and applicable candidates in OICs.

\begin{backmatter}
\bmsection{Acknowledgments} This work was supported by MSIT(Ministry of Science and ICT), Korea, under the ICT Creative Consilience program(IITP-2020-2011-1-00783) supervised by the IITP(Institute for Information \& communications Technology Planning \& Evaluation).
\end{backmatter}

\bibliography{References_OL}

\end{document}